# Landau Quantization and Highly Mobile Fermions in an Insulator

Pengjie Wang[1,#], Guo Yu[1,2,#], Yanyu Jia[1,#], Michael Onyszczak[1], F. Alexandre Cevallos[3], Shiming Lei[3], Sebastian Klemenz[3], Kenji Watanabe[4], Takashi Taniguchi[5], Robert J. Cava[3], Leslie M. Schoop[3], Sanfeng Wu[1,*]

[1] Department of Physics, Princeton University, Princeton, New Jersey 08544, USA
[2] Department of Electrical Engineering, Princeton University, Princeton, New Jersey 08544, USA
[3] Department of Chemistry, Princeton University, Princeton, New Jersey 08544, USA
[4] Research Center for Functional Materials, National Institute for Materials Science, 1-1 Namiki, Tsukuba 305-0044, Japan
[5] International Center for Materials Nanoarchitectonics, National Institute for Materials Science, 1-1 Namiki, Tsukuba 305-0044, Japan

[#] These authors contributed equally to this work
[*] Email: sanfengw@princeton.edu

**Abstract**

In strongly correlated materials, quasiparticle excitations can carry fractional quantum numbers. An intriguing possibility is the formation of fractionalized, charge-neutral fermions, e.g., spinons[1] and fermionic excitons[2,3], that result in neutral Fermi surfaces and Landau quantization[4,5] in an insulator. While previous experiments in quantum spin liquids[1], topological Kondo insulators[6–8], and quantum Hall systems[3,9] have hinted at charge-neutral Fermi surfaces, evidence for their existence remains far from conclusive. Here we report experimental observation of Landau quantization in a two dimensional (2D) insulator, i.e., monolayer tungsten ditelluride (WTe$_2$), a large gap topological insulator[10–13]. Using a detection scheme that avoids edge contributions, we uncover strikingly large quantum oscillations in the monolayer insulator's magnetoresistance, with an onset field as small as ~ 0.5 tesla. Despite the huge resistance, the oscillation profile, which exhibits many periods, mimics the Shubnikov–de Haas oscillations in metals. Remarkably, at ultralow temperatures the observed oscillations evolve into discrete peaks near 1.6 tesla, above which the Landau quantized regime is fully developed. Such a low onset field of quantization is comparable to high-mobility conventional two-dimensional electron gases. Our experiments call for further investigation of the highly unusual ground state of the WTe$_2$ monolayer. This includes the influence of device components and the possible existence of mobile fermions and charge-neutral Fermi surfaces inside its insulating gap.

**Main**

Bulk tungsten ditelluride (WTe$_2$) is a compensated semimetal in which an equal number of electrons and holes co-exist[14]. The semimetallic behavior remains when the material is thinned down to the trilayers[11,15]. In bilayers and monolayers, nevertheless, an insulating gap is observed[11], giving rise to the high-temperature quantum spin Hall effect in monolayers[10–13]. However, the mechanism for the gap opening remains mysterious[10,11,16,17]. The observation of superconductivity when the monolayer is doped with a low electron density[18,19] highlights the unusual nature of the insulating state.

**Device Design and the Insulator State**

The design of our devices for investigating the insulating state of monolayer WTe$_2$ is illustrated in Fig. 1a. A key here is to only detect the bulk conductance of the monolayer by avoiding contributions from the edge states. This is achieved by inserting a thin insulating layer of hexagonal boron nitride (hBN) between the palladium electrodes and the monolayer; selected small areas of the thin hBN were etched away so that only the tips of electrodes are in contact with the bulk without touching the edges. We employ a bottom contact geometry, which allows for the preparation of pre-patterned electrodes prior to the monolayer exfoliation, minimizing the flake's degradation. In a completed device, monolayer WTe$_2$ is encapsulated with top and bottom hBN layers, which also serve as gate dielectrics. An optical image of a typical device is shown as the inset of Fig. 1a. Detailed fabrication procedures are described in Methods, Extended Data Fig. 1 and Supplementary Information S1 and S2.

Figure 1b plots the two-probe resistance $R$ measured from device #1 (high impedance mode) as a function of the top gate voltage $V_{tg}$ while the bottom gate is grounded ($V_{bg} = 0$ V), at various temperatures $T$. The data were collected between two contacts separated by a distance of ~ 2 μm. Near zero gate voltage, a strongly insulating state develops with a low $T$ resistance of ~120 MΩ. This huge resistance indicates a successful elimination of the edge state contributions[11,13]. By either warming up the sample to near room temperature or applying a small gate voltage ($V_{tg}$ ~ 2.8 V), a decrease of four orders of magnitude in $R$ is observed, demonstrating the strongly insulating nature of the monolayer in the undoped regime. To better characterize the insulator, we performed four-probe resistance measurements (Fig. 1c). The resistance, $R_{4p}$, plotted in an Arrhenius scale at various $V_{tg}$, is generally captured by two regimes separated by a transition temperature near 100 K. By fitting the curve at $V_{tg}$ ~ -0.33 V to the activation model, $R_{4p}(T) \propto \exp(\Delta/2k_BT)$, where $k_B$ is the Boltzmann constant, we extract the gap $\Delta$ ~ 12.8 meV in the low $T$ regime and ~ 60.0 meV in the high $T$ regime, respectively. The size of the gap is controlled by $V_{tg}$ (Fig. 1c), as expected.

**Quantum Oscillations in the Insulator**

Surprisingly, the magnetoresistance of this highly insulating phase exhibits strikingly large quantum oscillations (QOs). Figure 2a plots $R$ as a function of magnetic field $B$ applied perpendicular to the two-dimensional (2D) atomic plane, at $V_{tg} = -0.33$ V and $T = 1.8$ K. On top of the large background resistance, the oscillation starts to develop with an onset field, $B_{onset}$, as small as ~ 1.5 T. The oscillation component $R_{osc}$ grows with increasing $B$, with an amplitude reaching ~ 60 MΩ near 8 T, about 2,300 times larger than the resistance quantum, $h/e^2$. Despite being such a strong insulator, the oscillation profile mimics the Shubnikov–de Haas (SdH) oscillations in a good metal[20]. The QOs have been robustly seen in additional devices (Extended Data Fig.2 for devices #2 and Extended Data Fig. 3 & 4 for device #3), different contact configurations (Extended Data Fig. 5) as well as dc measurements (Extended Data Fig. 6). No QOs are observed under in-plane magnetic fields (Extended Data Fig. 7). Figure 2b plots the oscillations as a function of $1/B$, which shows a clear SdH-type periodicity. Applying a fast Fourier transform (FFT) on the curve yields a sharp peak $\alpha$ at a frequency of ~ 48.6 T (Fig. 2c). If we adapt the standard analysis for SdH oscillations, this frequency indicates a quasiparticle density $n$ ~ 2.4×10$^{12}$ cm$^{-2}$, assuming a band with spin degeneracy. The oscillation profile, with many periods at a very small onset field, implies that the quasiparticles, which need to be fermions in order to give rise to the QOs, are highly mobile, with a mobility on the order of $1/B_{onset} \sim 6000$ cm$^2$V$^{-1}$s$^{-1}$.

To further characterize the QOs, we study the temperature dependence of the QOs. Warming the device up to $\sim 30$ K washes out the oscillations (Fig. 2b) in device #1. In Fig. 2d we plot the normalized oscillation amplitude $\Delta R/R_0$ as a function of $T$, where $R_0$ is the zero-field resistance and $\Delta R$ is the peak-to-valley difference for a given oscillation period at selected $B$ (Supplementary Information S3). The curves deviate from the standard Lifshitz-Kosevich (LK) description[20] (dotted line) and display a saturation at low temperatures. The two-component LK formula, which describes contributions from two Fermi pockets, also fails to fit our data (dashed line). However, the data is well captured by a modified LK formula, $\Delta R/R_0 \sim \left[\frac{1}{\gamma(B)} + \frac{\sinh(\beta T m^*/B)}{e^{-D/B}\beta T m^*/B}\right]^{-1}$, where $\beta = 2\pi^2 k_B/e\hbar$, $\hbar$ is the reduced Plank constant; $m^*$ is the effective mass of the carriers; $D$ describes the Dingle damping term and is related to the carrier mobility; and $\gamma(B)$ is a new term we introduce to describe the saturating behavior at low $T$. Here we ascribe the fitting parameters to physical quantities according to the standard analysis on SdH oscillations in metals. The fits (solid lines in Fig. 2d; see also Supplementary Information S3) yield $\gamma(B) \sim B^v$ with the exponent $v \sim 3.0$, and a $B$-dependent effective mass $m^*$ (see inset). The corresponding cyclotron energy is estimated to be $\sim 4.7$ meV near the highest field in our measurements (9 T), where $m^* \sim 0.22\, m_e$ ($m_e$ is the free electron mass). At $B_{onset} \sim 1.5$ T, the cyclotron energy will be even much smaller, and not comparable to the activation gap (Fig. 1c). The extracted quantum mobility $\mu$ is $\sim 1,100$ cm$^2$V$^{-1}$s$^{-1}$, a high value as expected. The results indicate the possible existence of highly mobile fermions in the insulator. If these fermions would be electrons or holes, a highly conductive sample would be expected, contradicting the observed strongly insulating behavior.

This conflict can be reconciled if the fermions are assumed to be charge neutral. Mobile neutral fermions can appear due to electron fractionalization[1–3,21]. Indeed, the search for fractionalized Fermi surfaces occupied by neutral fermions has been a major experimental quest in strongly correlated quantum matter, such as quantum spin liquids[1,4,22,23]. Theoretically, despite being charge-neutral, these fermions, e.g., spinons in spin liquids, may still couple to external magnetic fields through an internal gauge field and hence produce Landau quantization[4]. This concept has recently been extended to a different system, the mixed-valence insulators, in which a composite exciton Fermi liquid is proposed to exist[2,5]. QOs in resistivity are predicted to be observable on top of the activated charge transport[5], due to the Ioffe-Larkin rule[24]. Interestingly, this theory indeed predicts the low-$T$ saturation, which can be understood as a natural consequence of the unique $B$-dependent energy profile of neutral fermions[5].

Experimentally, evidence for QOs in insulators was reported only very recently in topological Kondo insulators under high magnetic fields[6,7] and InAs/GaSb quantum wells[25,26]. In contrast to our experiment, either no evidence for high mobility was extracted [6,7,25] or the sample remained conducting[26] (i.e., resistance $\sim h/e^2$) in these previous experiments. Consequently, multiple theories based on charge carriers[27–34], i.e., through Landau quantization of electronic bands outside the insulating gap, have been developed to capture these experiments without involving neutral fermions. Among them, several proposals predict QOs to appear in magnetization but not in charge transport[29–33], and hence they are not compatible with our observations. Thermal activation has been discussed to provide a channel for QOs to appear in resistivity[25,27,28,34]. However, neither do we observe the proposed activated behavior in QOs nor are such activated charge carriers in insulators expected to exhibit high mobility. Thus, our results are inconsistent with all existing scenarios based on charge carriers (see more discussions in Methods). More importantly, if the monolayer insulator indeed hosts highly mobile fermions, it may offer an unprecedented opportunity to examine the fully Landau quantized regime in an insulator, a completely new territory that was even unexplored in theories, including those involving neutral fermions.

**The Quantized Regime**

In Fig. 3, we summarize the key observations in another device (#2), where we indeed observed a quantized regime. Figure 3a plots two-probe resistance $R$ as a function of $V_{bg}$, taken at 2 K and 500 mK, respectively. To provide better contact, $V_{tg}$ is fixed at 2.5 V in this device. The highly insulating regime ($R > 1$ MΩ) is clearly developed when $V_{bg}$ is less than -1.8 V. Compared to device #1, the measured resistance strongly fluctuates and has a lower value, indicating that additional residual charge carriers exist in this device, likely due to disorders and inhomogeneities. This is consistent with a large magnetoresistance seen at small fields (< 1 T) in Fig. 3b (device #2) but absent in Fig. 2a (device #1), suggesting that these additional charge carriers in device #2 are rapidly localized. The sample also becomes as strongly insulating as device #1 when the field is tuned on. At 2 K, QOs appear at a similar $B_{onset}$ ~1.8 T (Fig. 3b), indicating a similar mobility compared to device #1. Here in Fig. 3b and c, the conductance $G$, defined as $1/R$, is plotted. Despite a lower oscillation frequency (~ 23 T, Fig. 3b inset), the QOs here show consistent results with device #1, as summarized in Extended Data Fig. 2 and 8.

When the device is cooled down to 500 mK, $B_{onset}$ shifts to an even smaller value, ~ 0.5 T (Fig. 3b inset), suggesting that the mobility is now much higher. Remarkably, at this temperature, the QOs develop into discrete peaks near 1.6 T, above which a quantized regime is fully developed (Fig. 3c). The data indicate that discrete Landau levels may be present inside the charge gap of the insulator. The low onset field of quantization and the evolving profile mimics the high-mobility conventional two-dimensional electron gases (2DEG). Conventionally, the conductance peaks in the quantized regime are associated with individual Landau levels at the Fermi level. If the Landau level were occupied by charge carriers, e.g., electrons or holes, the conductivity of the sample would be no different from other 2DEGs at this condition, i.e., the sample would be highly conducting at the peaks. However, the monolayer remains a strong charge insulator at all fields and the measured peak resistances remain on the order of ~ 100 MΩ (Fig. 3c). Our experiments here clearly rule out existing QO theories based on charge carriers and favor the presence of neutral fermions in the insulator.

**The Effect of Gate Voltages**

We next discuss gate-dependent studies, which provide additional insight on the observations. We note that the mobility of charge carriers in monolayer WTe$_2$ devices are typically low and as a result no QOs were observed in previous reports[11,13,18,19] even in the metallic regime. Indeed, Figs. 4a and b plot the magnetoresistance of devices #1 and #2, respectively, when the monolayer is electron doped, showing no QOs. Fig. 4c present a gate tuned color-map plot of $R_{osc}/R_0$ taken in device #1. While strong oscillations appear in the insulating plateau, their amplitudes quickly diminish when the monolayer is doped with either holes or electrons. A careful examination (Extended Data Fig. 9) reveals hints of QOs towards the electron-doped regime but it is very weak and only appears at the lowest measured temperature (1.8 K) in this device. In these devices with low charge mobilities, the QOs in the insulator are however very robust as previously presented. To enhance the visibility of the metallic QOs, in device #3, we have optimized our fabrication procedure including using a flux-grown bulk $WTe_2$ crystal of an exceptional quality. In this device, we achieved a significant improvement on the mobilities of charge carriers and hence can clearly resolve QOs not only in the insulating regime but also in the metallic regime. We find very rich behaviors of multiple QO branches and clear correlations between the metallic and insulating branches. A systematic analysis of the observations in device #3 is presented in the Methods section and Extended Data Fig. 3 and 4. While a comprehensive understanding of these observations is currently lacking, in all devices we find that the mobility of gate-induced charge carriers is decoupled with the robustness of the QOs in the insulating regime.

**The Scenario of Neutral Fermions**

Our experiments identify a variety of phenomena, in both the QO and fully quantized regimes, that point to the possible existence of highly mobile neutral fermions and neutral Fermi surfaces in the monolayer insulator. At this moment, the theory for such a neutral fermion picture in monolayer $WTe_2$ is yet to be developed. We mention two possible scenarios: (1) Since the insulating state can be converted to a superconducting state with a low electron doping, proposals based on neutral Majorana fermions in failed superconductors[21,35] should be examined; (2) It is likely that a mechanism analogous to the composite exciton Fermi liquid[2,3,5] is applicable to monolayer $WTe_2$. In this scenario, electrons coexist with an equal number of holes, which, due to correlations, split into charged bosonic holons and charge-neutral fermionic spinons (spin-charge separation). Coulomb attraction binds electrons and holons to form fermionic excitons, leaving behind a charge gap and mobile neutral fermions, e.g., spinons and fermionic excitons[2]. Note that in the semimetallic $WTe_2$ trilayers QOs reveal that electrons and holes coexist at similar densities[15], and early calculations on the monolayer also predict the coexistence of electrons and holes[10]. In a separate manuscript[36], we present experimental evidence that supports the presence of an excitonic insulator phase in the monolayer based on a systematic study combining transport and tunneling measurements as well as theoretical modeling. We note that the reduced dimensionality and in-plane anisotropy, as well as the monolayer lattice being a distorted version of the dice lattice[37], could set the stage for strong correlations and the charge-statistic separation.

**Conclusion**

The results call for future investigations on the highly unusual insulating ground state of monolayer $WTe_2$ in both theory and experiments, and encourage the search for fractionalized, neutral Fermi surfaces inside its charge insulating gap. Further explorations in the quantized regime promise key advances in understanding the exact nature of the insulator. The connections between the possible fractionalized insulating phase and superconductivity, as well as nontrivial topology, deserve a careful study, for which monolayer $WTe_2$ provides an excellent platform. The experimental search for quantum phases of neutral fermions, even beyond the Fermi surface state, will lead to a new chapter in quantum matter.

## Acknowledgements

S.W. is indebted to N. P. Ong and A. Yazdani for their support to his new lab and their encouragement and discussions regarding this work. We acknowledge discussions with B. A. Bernevig, D. Cobden, F. D. M. Haldane, P. Jarillo-Herrero, P. A. Lee, T. Senthil, I. Sodemann, S. Sondhi, and X. Xu. This research was supported by NSF through a CAREER award to S. Wu (DMR-1942942) and the Princeton University Materials Research Science and Engineering Center (DMR-1420541). Early measurements were performed at the National High Magnetic Field Laboratory, which is supported by NSF Cooperative Agreement No. DMR-1644779 and the State of Florida. K.W. and T.T. acknowledge support from the Elemental Strategy Initiative conducted by the MEXT, Japan, Grant Number JPMXP0112101001, JSPS KAKENHI Grant Number JP20H00354 and the CREST(JPMJCR15F3), JST. F.A.C. and R.J.C. acknowledge support from the ARO MURI on Topological Insulators (grant W911NF1210461). S.L, S.K., and L.M.S. acknowledge support from the Arnold and Mabel Beckman Foundation through as Beckman Young Investigator grant awarded to L.M.S and the Gordon and Betty Moore Foundation through Grant GBMF9064 to L.M.S.

## Author Contributions

P.W., G.Y., Y.J., and M.O. fabricated devices and built measurement systems. P.W., G.Y., Y.J., and S.W. performed measurements and analyzed data. S.W. designed and supervised the project. F.A.C., R.J.C., S.L., S.K, and L.M.S. grew and characterized bulk WTe$_2$ crystals. K.W. and T.T. provided hBN crystals. S.W. and P.W. wrote the paper with input from all authors.

## Competing Interests

The authors declare no competing financial interests.

## Data Availability

The data that support the findings of this study are available from the corresponding author upon reasonable request.

## Methods

### Sample Fabrication

The WTe$_2$ bulk crystals were grown using similar methods described in previous works[14,38]. The exfoliation and search of hBN and graphite flakes were performed under ambient conditions. The sample contacts were patterned by electron beam lithography (EBL), followed by a cold develop, reactive ion etching (RIE), and metal deposition (3 nm Ti/17 nm Pd). The holes on the thin hBN for contacts were patterned by EBL and etched by RIE. The exfoliation, search, and transfer of monolayer WTe$_2$ flakes were performed in a glovebox equipped with a dry transfer setup. Step-by-step sample fabrication procedures are illustrated in Extended Data Fig.1 and Supplementary Information S1.

### Transport Measurement

The electrical measurements of our devices were performed either in a cryostat (Quantum Design Dynacool) equipped with a superconducting magnet or in a dilution refrigerator (Bluefors LD250) equipped with a superconducting magnet. Standard lock-in measurements were taken with a low-frequency (< 7 Hz) ac excitation (< 5 mV). A current preamplifier (DL Instrument 1211) was used to improve signals.

**Contact Resistance**

Our key resistance data were taken in two-probe measurements, which involve contact resistance. We note that the presence of contact resistance does not alter our conclusions since contact resistance is not expected to give rise to QOs. A large contact resistance can only lower the visibility of the QOs (i.e., $R_{osc}/R$). Hence the fact that pronounced QOs are seen implies that the contact resistance is small compared to the sample resistance. The contact resistance may introduce a small correction to the LK fittings (Fig. 2d, Extended Data Fig. 8 and Fig. S1), but the only impact would be a slight change of the extracted parameters (e.g., the effective mass), which are not essential to our conclusions. One advantage of our device geometry is that the gates provide a useful knob to improve the contact properties if the pristine version is not good enough.

**Further Discussion on Excluding Trivial Explanations on the QOs**

Here we add further discussion on excluding trivial scenarios involving unexpected contaminations from any metallic regimes that may present in the devices. The metallic components in our devices include edge states of the monolayer, graphite gates, metal electrodes, and possible metallic islands in the monolayer due to impurities or bubbles, etc. We first note that our device is designed to avoid the edge contributions, which is the key to observe the QOs (absent in previous devices involving edges). In addition, there is no known mechanism for 1D helical edge modes to give rise to QOs with periodicity in $1/B$. Next, we present systematic investigations on alternative explanations based on graphite. (1) Our graphite gates are well separated from the WTe$_2$ channel by the hBN dielectric. Within our applied gate range, no measurable leak current presents (i.e., the resistance between graphite and WTe$_2$ is above our measurement limit, > 100 GΩ). However, the measured channel resistance in the QO regime is ~ 100 MΩ, suggesting that the current flow is restricted to the monolayer. The strict dc measurements, which show consistent QOs (Extended Data Fig. 6), indicate that the QO signal is unrelated to any unexpected capacitive coupling to the graphite. (2) The graphite is also not expected to develop QOs that are qualitatively similar to the ones observed in our devices, which are dominated by a single frequency (see a recent study[39], which shows that, QOs developed in thin graphite film, even if fully encapsulated by hBN from both top and bottom sides, is highly irregular below 10 T). Also, in our case, the thin graphite gates are not fully encapsulated and the graphite density of states are not expected to be fully gapped[39] in magnetic fields. (3) We have fabricated and measured three devices without using graphite gates（Extended Data Fig. 10). While the quality of these devices is significantly lower than the graphite-gated devices, hints on magnetoresistance oscillation with $1/B$ periodicity were seen in the insulating regime in the sub-Kevin regime. (4) Due to the capacitance effect between the two gates (especially when the monolayer WTe$_2$ is an insulator), a changing gate voltage (either the top or bottom gate) will inevitably tune the carrier density of both top and bottom graphite significantly. Hence the unchanged frequency of the QO branches observed in device #1 and #3 is inconsistent with assigning the QOs to the graphite layer. (5) The use of graphite gates has been widely used for 2D devices, e.g., graphene or transition metal dichalcogenides, where graphite QOs are invisible in the transport measurements of these devices. Especially in the metallic regime of device #3 (with doping well above the MIT), the situation is directly comparable to graphene devices. (6) The gate tuned QOs observed in device #3 exhibit very rich behaviors, including the appearance of multiple branches, the rich dependence on the displacement field, the clear correlations between metallic and insulating branches, and an intermediate regime where QOs are absent (see Methods section). It would be extremely unlikely to attribute all these behaviors to a graphite artifact with any comprehensive physical meaning. Thus, we conclude that the observed QOs arise due to the unusual intrinsic properties of the monolayer $WTe_2$.

In general, for any metallic channels to be responsible for the QOs, they cannot be in parallel with the monolayer insulator, since otherwise it will short the insulator and contradict the huge measured resistance. They cannot be in series with the monolayer insulator either, as the amplitude of SdH oscillations in metals is only expected to be on the order of $h/e^2$, which contradicts the large QO amplitude observed in our devices (e.g., $R_{osc}$ reaches ~ 60 MΩ in Fig. 2). QOs from a metallic regime in series are expected to be invisible on top of the huge background resistance. Moreover, in the quantized regime of device #2 (Fig. 3c), the zero conductance between peaks corresponds to a resistance value reaching ~ 100 GΩ, implying that the contributions from any metallic regime have been completely suppressed. Our data cannot be explained by the presence of any metallic component listed above.

**Analysis of Data Taken from Device #3 (with high mobility of charge carriers)**

In the main text, we have presented gate-dependent studies of the QO signals observed in device #1, where we have observed a dominant mode with a frequency at ~ 48.6 T and a slow modulation at ~ 13.2 T in the insulating state. A very weak branch appears towards electron doping only at the lowest measured temperature (1.8 K). The weakness (device #1) or absence (device #2) of QOs in the metallic regime indicates that the mobilities of charged carriers are low. In device #3 (see device image in Extended Data Fig. 3a), we have significantly improved the mobilities of charge carriers, by using an optimized fabrication method that minimizes disorder and a much-improved flux-grown $WTe_2$ bulk crystal (with an exceptional residual resistivity ratio ~ 2500). This improvement allows us to clearly resolve QOs not only in the insulator regime but also in the metallic regime.

We first compare device qualities in Extended Data Fig. 3b, by showing the resistance of the three devices (#1-3) as a function of the gate induced carrier density $n_g \equiv \varepsilon_r\varepsilon_0(V_{tg}/d_{tg}+V_{bg}/d_{bg})/e$, where $e$ is the elementary charge, $\varepsilon_r$ is the relative dielectric constant of hBN, $\varepsilon_0$ is the vacuum permittivity and $d_{tg}$ ($d_{bg}$) is the thickness of hBN layer associated with the top (bottom) gate. The data were taken at 70 K. One immediately sees that the resistance peak of device #3 is about an order of magnitude narrower than the other two, indicating significantly reduced disorders for charge carriers. The offset of the peak maximum (i.e., charge neutrality point, CNP) from zero $n_g$ is another impurity indicator as it reveals the unintentional doping during the fabrication process. This offset is also much reduced in device #3, again demonstrating its excellent quality. In Extended Data Fig. 3c, we show the temperature-dependent $R_{4p}$, revealing a metal-insulator transition (MIT) that occurs at $n_g \sim 2.4 \times 10^{12}$ cm$^{-2}$. In a separate manuscript[36], we present systematic studies on the transport properties of this device, including the dual-gated resistance map and Hall response at a relatively high-temperature regime. The results there support the insulating state of monolayer WTe$_2$ as an excitonic insulator and rule out alternative explanations such as a band insulator or a localized insulator[36]. Here we focus on the QOs, which appear at low temperatures.

Extended Data Figs. 3e and f illustrate the QO frequency map under varying $n_g$ at a selected displacement field $D$ ($D \equiv (V_{bg}/d_{bg} - V_{tg}/d_{tg})\varepsilon_r/2$). The corresponding traces of the scanning $n_g$ are illustrated in Extended Data Fig. 3d. The FFT amplitude is normalized to its individual maximum of each magnetoresistance curve at a given $n_g$ to enhance the visibility of QOs at all gates. The data was taken at 7 K to minimize the bad contact regime. The QO map with $D \sim 0.15$ V/nm (a value we believe is needed to cancel the residual displacement field in the devices, i.e., corresponding to zero net $D$; see more discussions in ref. [36]) clearly reveals an electron-like ($e$-like) branch in the electron-doped metallic regime and a hole-like ($h$-like) branch in the hole-doped regime (Extended Data Fig. 3e). Interestingly, near the CNP, an additional branch, labeled as $\alpha_1$, develops at a frequency $\sim 17$ T. The appearance of this $\alpha_1$ peak is consistent with the observation in device #1 (Extended Data Fig. 9). We note that while the flake quality is significantly improved here, the contact properties in the insulating regime are however worse than device #1. Device #1 so far exhibits the best contact properties among all fabricated devices, which allow us to observe QOs without applying a top gate voltage (i.e., at CNP near $D = 0$, Extended Data Fig. 9). With $D = 0.59$ V/nm, more features develop as shown in Extended Data Fig. 3f. In the electron-doped regime, one additional $h$-like branch appears accompanied with an increased frequency of the $e$-like branch. In the insulating state near CNP, we now see two branches ($\alpha_2$ and $\beta$). One branch, labeled as $\alpha_2$, appears at a very close frequency of $\alpha_1$. The $\beta$ branch appears at a location that is strongly correlated with the $e$- and $h$-like branches in the doped regime. We find that the $\beta$ branch consistently shows up at the intersection of the $e$- and $h$-like traces if one extends them into the insulating regime, i.e., the three branches ($e$, $h$, and $\beta$) form a tilted "Y" shape (Extended Data Fig. 3g). The effect of displacement field will be further discussed below. The observation suggests that the insulating $\beta$ branch shares the same origin as the $e$- and $h$-like branches.

While the frequency of the $\beta$ peak is gate-tunable, we next carefully examine the insulating $\alpha_2$ branch and show that its frequency is gate independent. A careful examination on the $h$-like branch shown in Extended Data Fig. 3f reveals an interesting wiggling behavior near the frequency of $\alpha_2$. This is due to the presence of an additional branch ($\alpha_3$) in the metallic regime. To confirm its existence, we plot the QO frequency map under varying the displacement field $D$ at a fixed $n_g = 2 \times 10^{12}$ cm$^{-2}$. We observe that frequencies of the $e$- and $h$-like branches increases with increasing $D$. This displacement field effect is consistent with the expectation that the application of $D$ to WTe$_2$ monolayer enlarges the Fermi surface of both electrons and holes without altering the total charge density (see more experimental and theoretical discussions in ref.[36]). Clearly, in addition to the $e$- and $h$-like branches, branches located near the $\alpha_2$ frequency develop in the map (Extended Data Fig. 4a). We have identified these $\alpha$ branches by labeling them as $\alpha_3$ - $\alpha_6$. Their emergence appears to be correlated with the $e$- or $h$-like branches. For instance, it looks like that the $\alpha_3$ and $\alpha_6$ branches may arise due to a splitting associated with the $h$-like branch as one increases $D$.

Interestingly, once the $\alpha$ branches develop, the gate voltages have only very weak or no impact on their frequency. To further reveal the gate dependence of the $\alpha_3$ branch, we plot its behavior under varying a single gate ($V_{tg}$ or $V_{bg}$) while fixing the other (Extended Data Fig. 4b and c). It is clear that both gates are not able to tune the frequency of the $\alpha_3$ branch. This $\alpha_3$ branch is in fact the one that produces the wiggling behavior in the $h$-like branch in Extended Data Fig. 3g, hence correlated to the insulating $\alpha_2$ branch. To examine the gate dependence of the $\alpha_2$ branch, in Extended Data Fig. 4f, we fix the charge density at CNP and vary the displacement field $D$, under which both gates are varied toward opposite directions. The map clearly reveals a gate-tunable $\beta$ branch and a gate-independent $\alpha_2$ branch. In Extended Data Fig. 4g, we summarize the appearance of the $\alpha$ branches in the measured maps.

While a comprehensive understanding of the existence and gate dependence of these QO branches is currently lacking, we make several comments on the observations. (1) The observation of the gate-independent $\alpha$ branch here is consistent with device #1 (Extended Data Fig. 9), where the observed insulating branches (despite the presence of two frequencies) show no dependence on the applied top gate voltage. (2) One important difference between the two devices is the measurement regime in the parameter space spanned by $n_g$ and $D$. Device #1 exhibited the best contact properties among all of our fabricated devices and it is so far the only device that allows us to perform measurements in the insulating regime without applying a top-gate (e.g., $V_{tg} = 0$ V, or near $D = 0$ at CNP, Fig. 4c). In all other devices including device #3 shown here, we have to apply a finite $V_{tg}$ in order to achieve good contacts. Namely, we do not have QO information around CNP near $D = 0$ for device #3. The contact issue needs to be considered when comparing different devices. (3) The appearance of both $e$- and $h$-like branches is remarkably consistent with the excitonic insulator picture for the insulating state (see ref. [36]). (4) We note that there is also an intermediate regime below the MIT where all QOs are absent (Extended Data Fig. 3g). This could indicate a nontrivial interacting effect of the correlated electron-hole system near the transition. (5) In Extended Data Figs. 4b and c, the $e$- and $h$- branches appear to be individually controlled by the top and bottom gates in the doped regime. This is consistent with the expectation that both the density tuning and the displacement field effect come into play with one tune a single gate. For example, increasing $V_{tg}$ will lead to an increasing magnitude of $D$ as well as an increasing $n_g$. If $D$ enlarges both $e$- and $h$- pockets (see discussions in ref[36]) while $n_g$ enlarges $e$-pocket yet reduces $h$-pocket, then a net result can be that only $e$-pocket size is enlarged. The effect can also be understood in real space, as shown in Extended Data Fig. 4e. With the applied gate voltages, the presence of a finite $D$ will polarize the electron and hole distributions in the out of plane direction (note that the WTe$_2$ monolayer is made of three atomic layers). In this case, the top gate couples to the electron layer while the bottom gate couples to the hole, naturally producing the gate dependence seen in Extended Data Fig. 4b and c. (6) We further emphasize that the gate response of an excitonic insulator is a rather complicated process, which is distinct from a simple system like graphene. For instance, in a correlated electron-hole system, the system can adjust either the hole or the electron concentrations in response to the same change of the total charge density. As shown in ref [40], even in a simplified model of an artificial electron-hole bilayer system, the gate response is highly nontrivial. The exact response of an excitonic insulator further depends on microscopic details including correlations and disorders. In our case, the electrons and holes are much more confined in a single WTe$_2$ layer, where strong electron-hole correlations are expected.

The highly unusual ground state of the monolayer WTe$_2$ (likely the excitonic insulating ground state) is the key to understand various observations presented here.

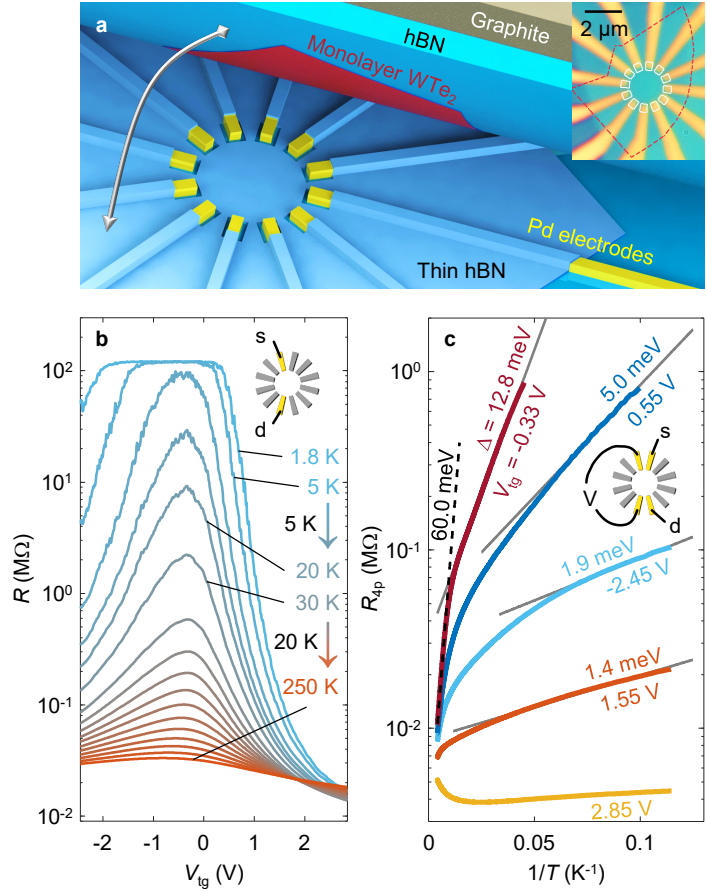

**Figure 1 | Device schematic and the insulating state of monolayer WTe$_2$. a.** Cartoon illustration of the device design, where a thin hBN layer with selectively etched areas is used to avoid contact to the edges of the monolayer WTe$_2$. The inset shows an image of device #1: the dashed red line highlights the monolayer edges and the white squares denote the contact regions. **b.** Gate-dependent two-probe resistance, taken from device #1, at various temperatures. The inset shows the measurement geometry (s: source; d: drain). **c.** Temperature dependence of four-probe resistance, measured with a geometry shown as an inset, at selected gate voltages, $V_{tg}$. Grey solid (black dashed) lines are the fits to the thermal activation model in the low (high) temperature regime. The extracted activation gaps and $V_{tg}$ are labeled next to the curves. Data taken from device #1.

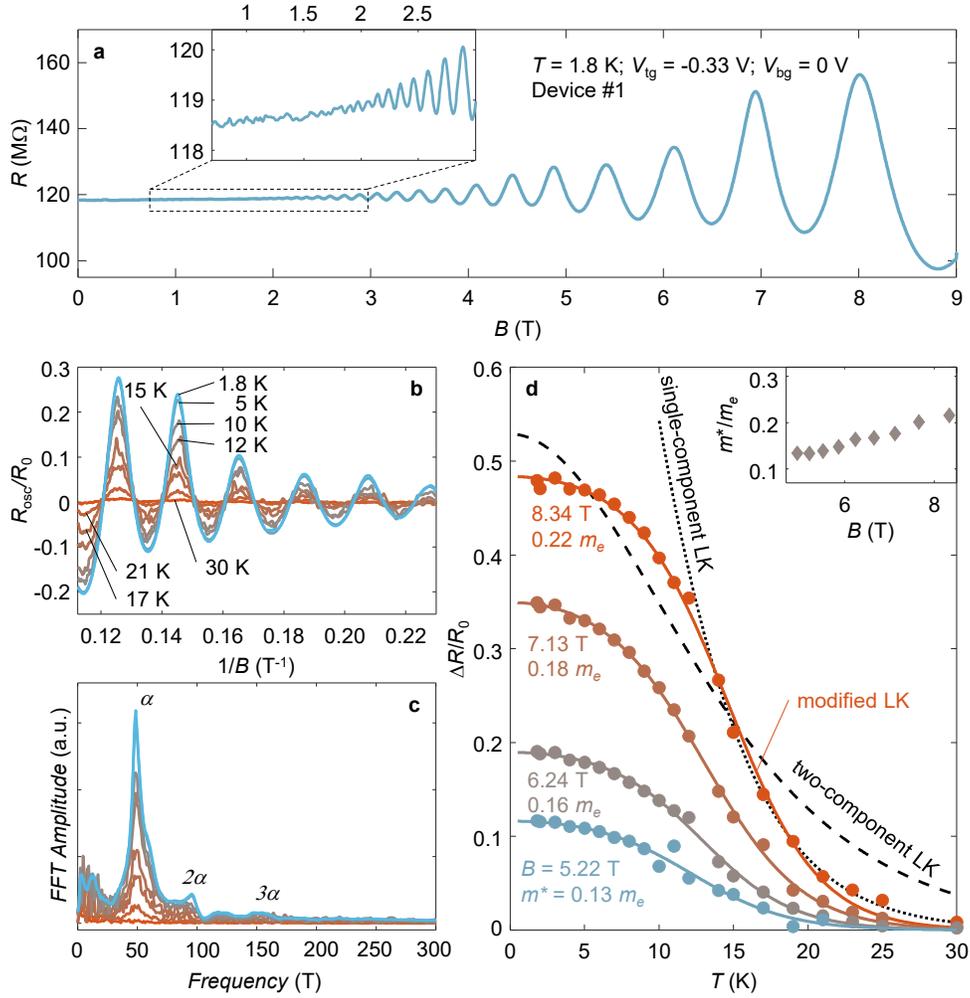

**Figure 2 | Quantum oscillations in the monolayer insulator. a.** A typical low-temperature magnetoresistance curve taken in the insulating plateau of device #1. The magnetic field $B$ is perpendicular to the 2D atomic plane. The inset zooms in on the low field data. **b.** Temperature dependence of the oscillating component $R_{osc}$, normalized by the zero-field resistance $R_0$, as a function of $1/B$. **c.** FFT on the same magnetoresistance data. **d.** LK fittings to the temperature-dependent oscillations at selected $B$. The inset plots the extracted effective mass. Details of the fitting procedures are described in the Supplementary Information S3.

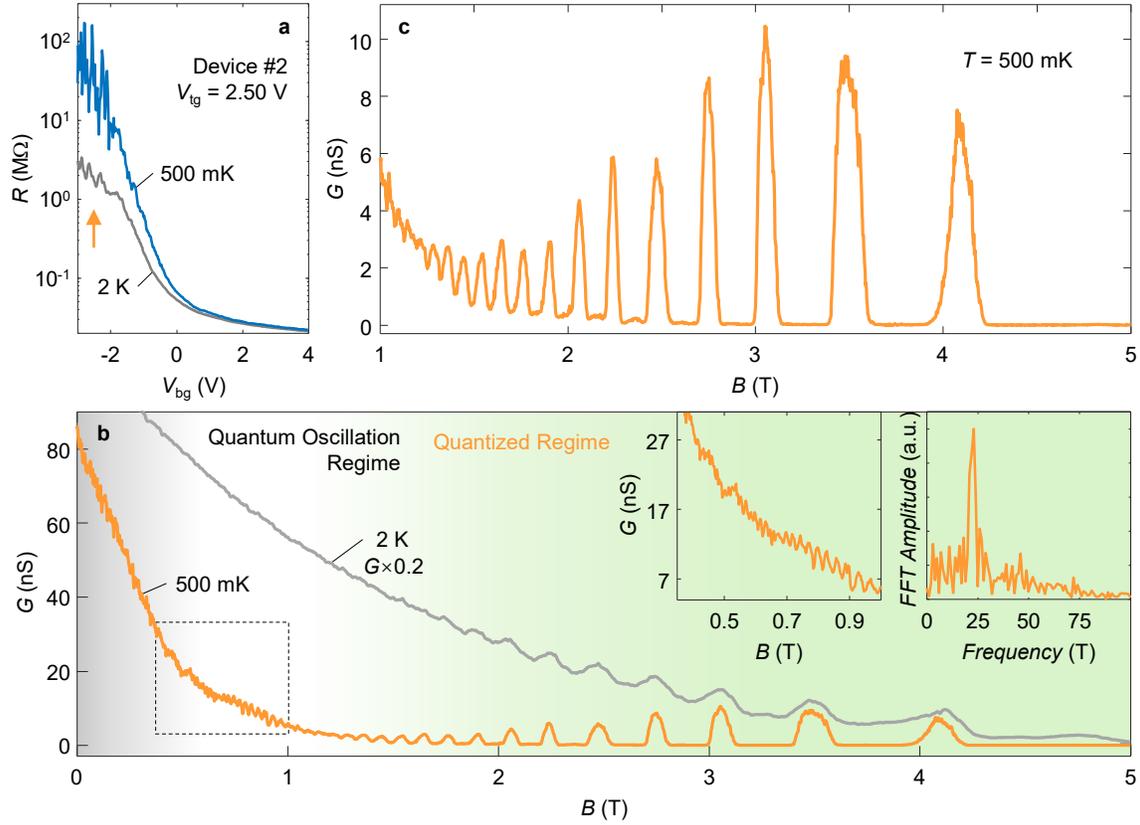

**Figure 3 | Quantized regime and discrete Landau levels in the monolayer insulator. a,** Two-probe resistance $R$, taken from device #2, as a function of $V_{bg}$ ($V_{tg}$ = 2.5 V). **b,** Magnetic field dependent conductance $G$, defined as $1/R$, taken at a selected gate voltage ($V_{bg}$ = - 2.5 V, indicated by the arrow in **a**). Data taken at 500 mK (orange) clearly displays the fully quantized regime, where discrete Landau levels are observed. Data at 2 K (grey) are shown as a reference. Left inset zooms in on the orange curve observed below 1 T, as indicated by the dashed box. Right inset shows FFT of the 500 mK data between 0.6 T to 1.5 T. **c,** Zoomed plot of the data above 1 T in **b**.

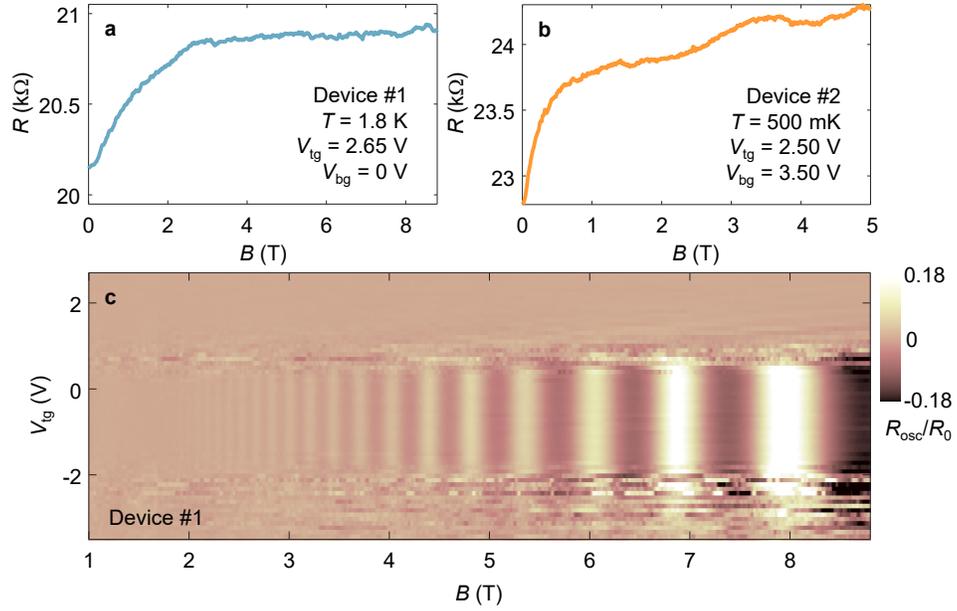

**Figure 4 | Gate dependence of the quantum oscillations. a,** A typical magnetoresistance curve measured from device #1 when the monolayer is electron-doped by gating. **b**, a similar plot to **a**, but taken from device #2. The slight wiggles are likely due to disorder. Both show the absence of QOs. **c**, Color map of the normalized oscillating component, $R_{osc}/R_0$, under varying $B$ and $V_{tg}$, taken from device #1 at 1.8 K.

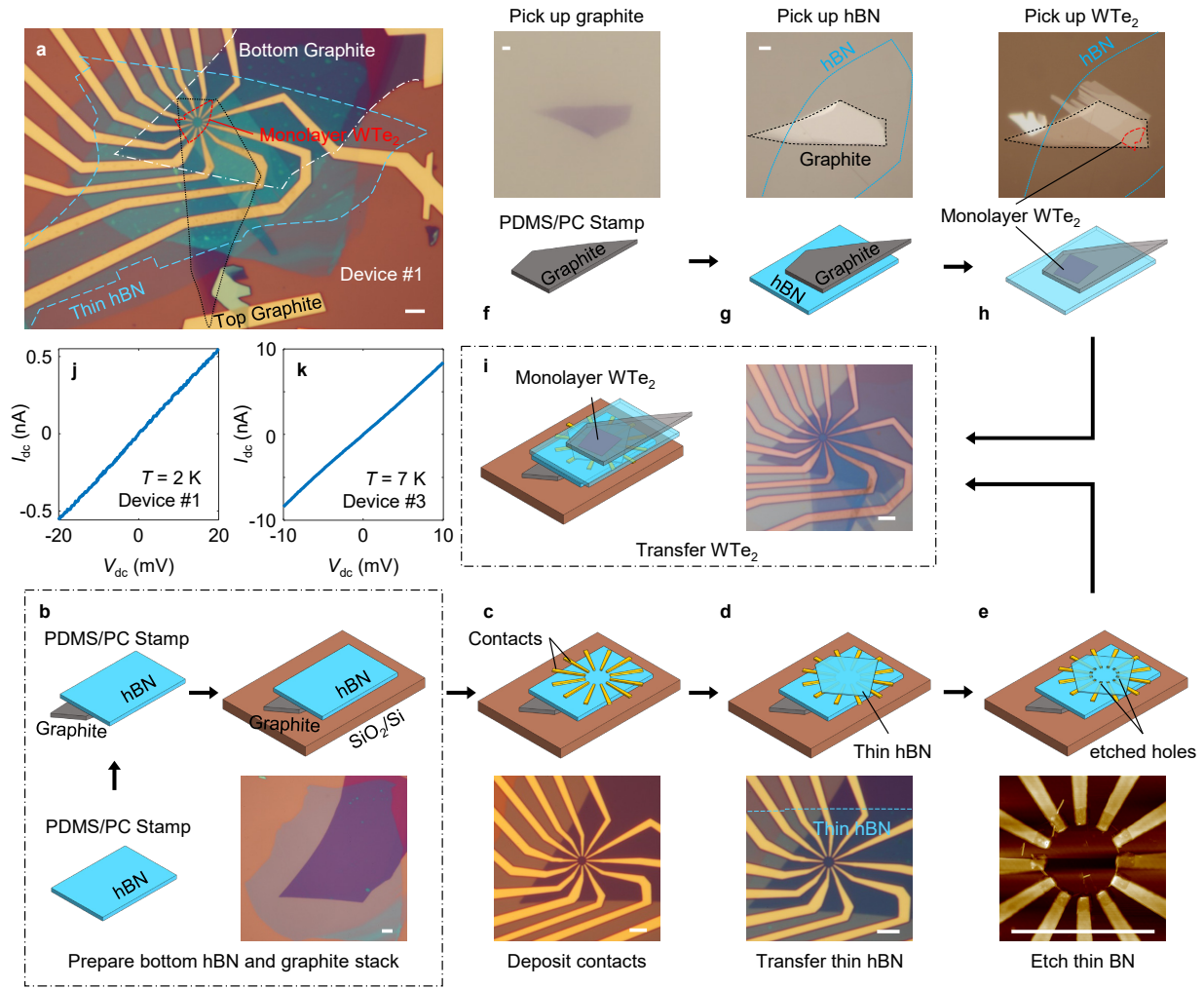

**Extended Data Fig. 1 | Sample fabrication process. a,** An optical image of device #1. The monolayer WTe$_2$ is highlighted by the red dashed line. **b,** Bottom hBN and graphite preparation. **c,** Deposition of metal contacts (3 nm Ti and 17 nm Pd). **d,** Transfer of thin hBN in a dry transfer system under the ambient condition. **e,** Etching thin hBN at the tips of the electrodes. **f,** Picking up top graphite. **g,** Picking up top hBN. **h,** Picking up monolayer WTe$_2$. Pictures in **g** and **h** are taken by flipping over the stamp after picking up the flakes. No visible bubbles are observed. **i,** Transfer monolayer WTe$_2$ to the bottom part prepared in **b-e**. **j and k** Typical *I-V* curves (dc measurement) of device #1 (#3) between two contacts at 1.8 K (7 K) showing ohmic behavior. All scale bars are 4 μm.

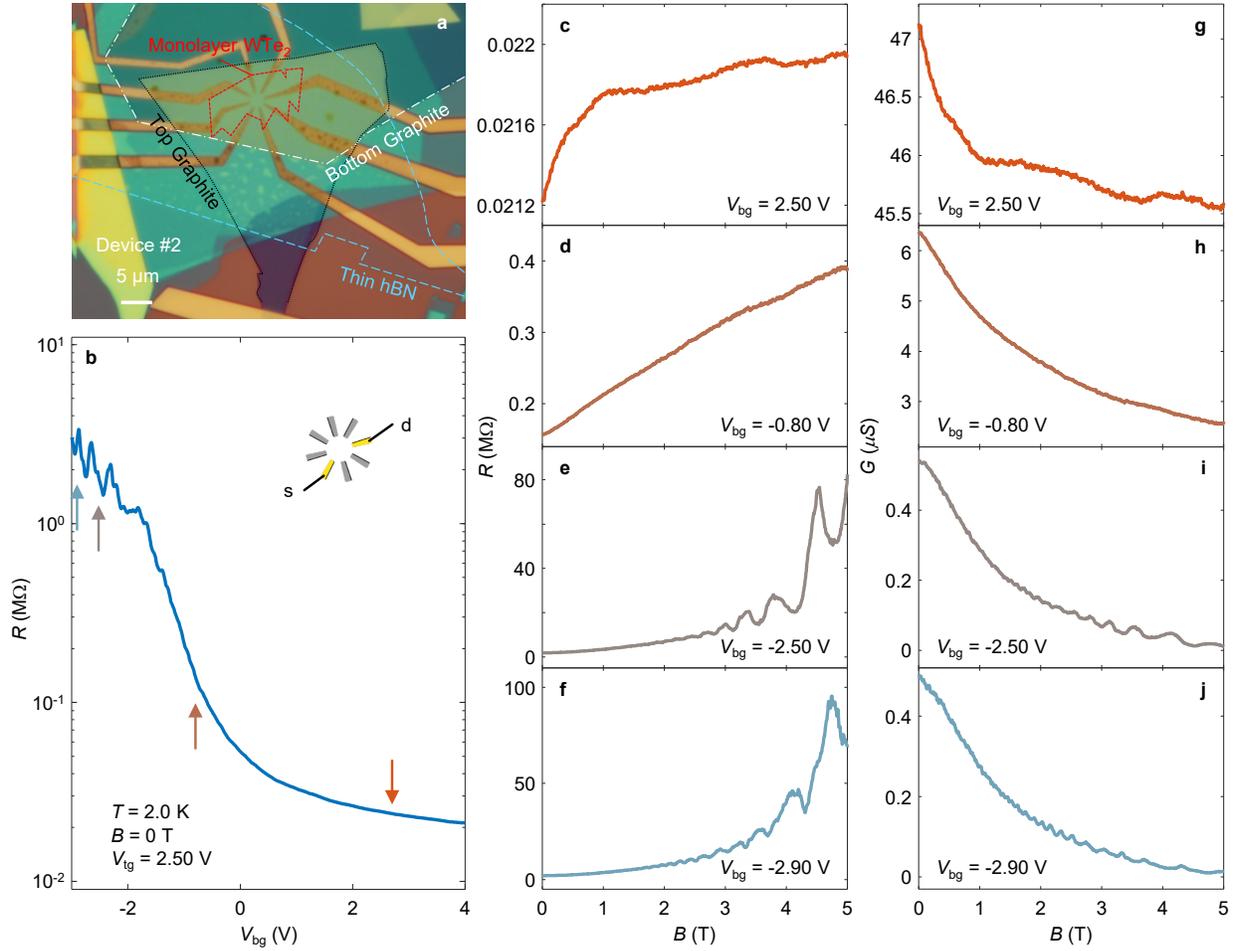

**Extended Data Fig. 2 | Summarized data from device #2. a,** An optical image of device #2. The red dashed line highlights monolayer WTe$_2$. **b,** Gate dependent two-probe resistance $R$. The inset shows the measurement geometry. The colored arrows indicate the gate voltages used in **c-f**. **c-f,** Raw data of the magnetoresistance curves taken at selected $V_{bg}$. Strong QOs are observed in the insulating regime but disappear when the monolayer is doped. **g-j,** The same data in **c-f**, but plotted in $G$.

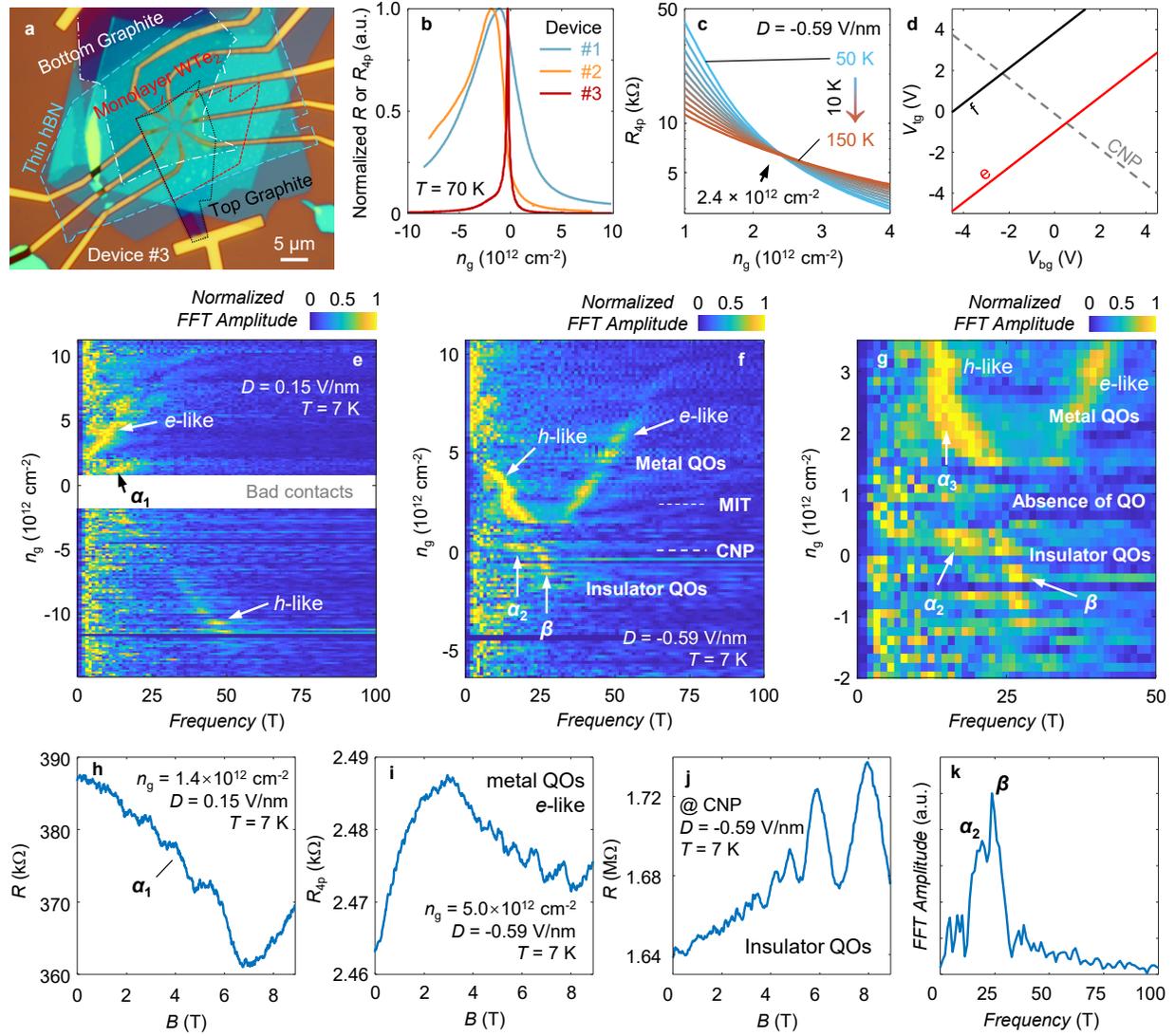

**Extended Data Fig. 3 | Quantum oscillations observed in device #3 (with high mobility for charge carriers) – part 1. a,** An optical image of device #3. The red dashed line highlights monolayer WTe$_2$. **b,** Comparison of the gate-tuned resistance of the three devices at 70 K (normalized $R_{4p}$ for device #2 & #3 and normalized $R$ for device #1). Here $n_g \equiv \varepsilon_r\varepsilon_0(V_{tg}/d_{tg}+V_{bg}/d_{bg})/e$, where $e$ is the elementary charge, $\varepsilon_0$ is the vacuum permittivity, $\varepsilon_r$ is the relative dielectric constant of hBN and $d_{tg}$ ($d_{bg}$) is the thickness of the hBN layer associated with the top (bottom) gate. **c,** Gate-tuned $R_{4p}$ of device #3 with a fixed $D = -0.59$ V/nm ($D \equiv (V_{bg}/d_{bg} - V_{tg}/d_{tg})\varepsilon_r/2$)), showing the metal-insulator transition (MIT). **d,** Schematic diagram for gate-tuning traces used in **e** (red) and **f** (black). The dashed line indicates CNP. **e,** Normalized FFT amplitude of the QOs under varying $n_g$ at $D = 0.15$ V/nm (a value that we believe is needed to cancel out the residual displacement field presented in the device, i.e., the net total $D$ here is expected to be zero). The FFT amplitude is individually normalized to its individual maximum for a given $n_g$ in order to enhance the visibility at all $n_g$. **f,** $n_g$-tuned FFT amplitude (normalized) of the QOs at $D = -0.59$ V/nm. QOs in both the metallic (*e*-like and *h*-like branches) and the insulating regime ($\alpha_2$ and $\beta$

branches) are clearly observed. **g,** Zoom-in plot of the FFT features near CNP in **f**, highlighting the correlations between the insulating and metallic branches. **h,** The magnetoresistance trace corresponding to the $\alpha_1$ branch in **e** ($n_g$ = 1.4 × $10^{12}$ cm² and $D$ = 0.15 V/nm). **i,** The magnetoresistance trace ($R_{4p}$) of the metallic QO branch in **f** ($n_g$ = 5.0 × $10^{12}$ cm² and $D$ = -0.59 V/nm). **j,** Magnetoresistance ($R$) trace taken at CNP with $D$ = 0.59 V/nm, showing QOs in the insulating phase. **k,** FFT of oscillation component in **j**. Discussions on these observations are presented in the Methods section.

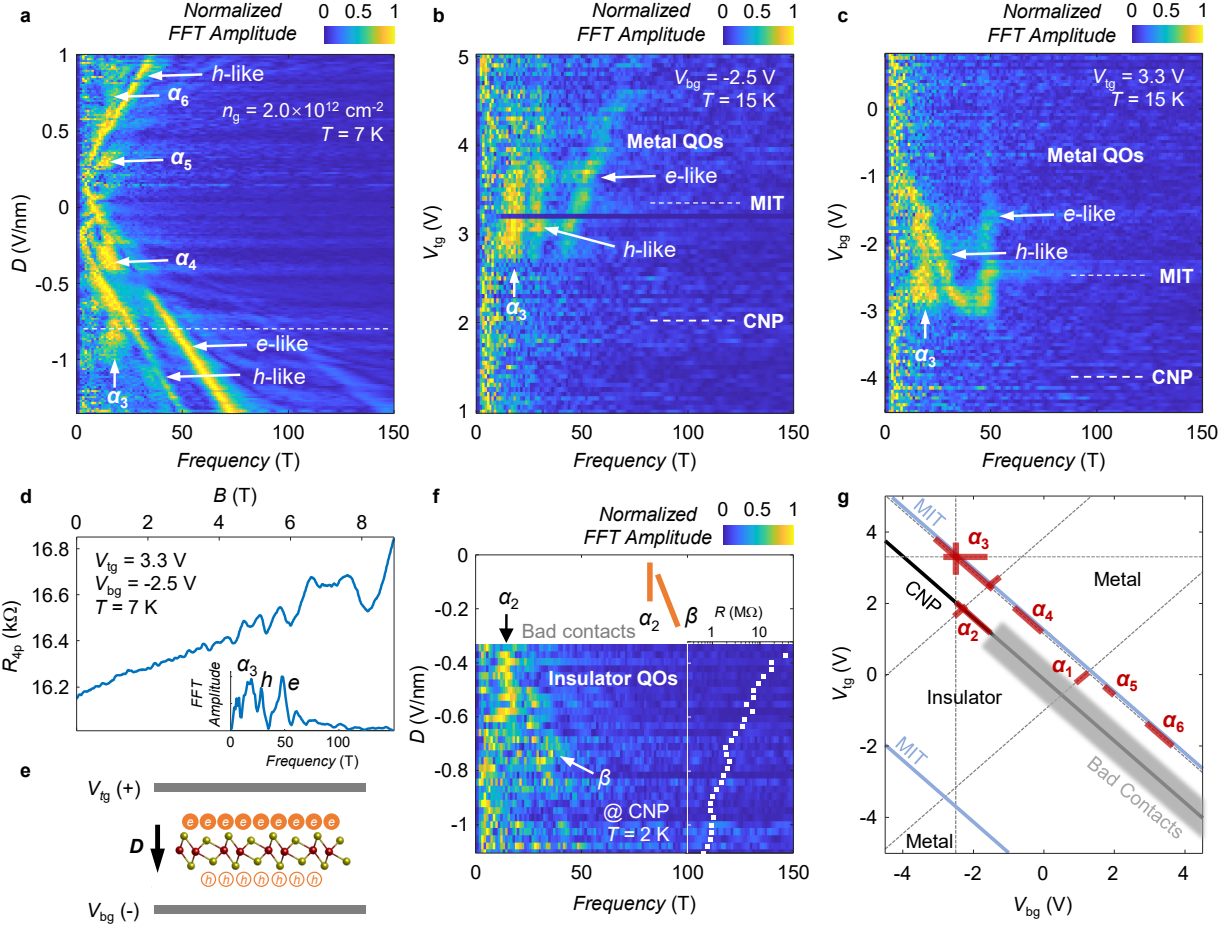

**Extended Data Fig. 4 | Quantum oscillations observed in device #3 (with high mobility for charge carriers) – part 2. a,** $D$-tuned QO frequency map (normalized) at $n_g = 2.0 \times 10^{12}$ cm$^2$. $T = 7$ K. Visible branches ($\alpha_3$-$\alpha_6$, $e$-like and $h$-like) are noted. **b,** $V_{tg}$-tuned QO frequency map at $V_{bg} = -2.5$ V and $T = 15$ K. **c,** $V_{bg}$-tuned QO frequency map at $V_{tg} = 3.3$ V and $T = 15$ K. Three visible branches of QOs in **b** and **c** are labeled as $\alpha_3$, $h$-like, and $e$-like. At $T = 15$ K, no QOs are observed in the insulator regime. **d,** Raw data of the QOs at $V_{tg} = 3.3$ V and $V_{bg} = -2.5$ V, corresponding to the dashed line in **a**. Inset plots the FFT of the oscillating amplitude, showing three peaks as labeled. **e,** Illustration of the monolayer WTe$_2$ with top and bottom gates geometry. The presence of the displacement field polarizes the electron distribution. This results in the separation of the electrons and holes, which can then couple to the top and bottom gates respectively. **f,** $D$-dependence of the insulator QOs ($\alpha_2$ and $\beta$ branches) at CNP ($T = 2$ K). Inset plots the corresponding resistance at zero $B$ (the resistance increases further with $B$), confirming the insulating nature. The behaviors of the $\alpha_2$ and $\beta$ branches are also sketched (solid orange lines). **g,** A sketch summarizing the appearance of the gate-independent $\alpha$ branches ($\alpha_1$-$\alpha_6$) on the maps that we have presented. The gray dashed lines are traces that were examined. The blue solid lines indicate the MIT. The black solid line indicates CNP. The gray area indicates the bad contact region, where information about QOs is inaccessible in our measurement. Discussions on these observations are presented in the Methods section.

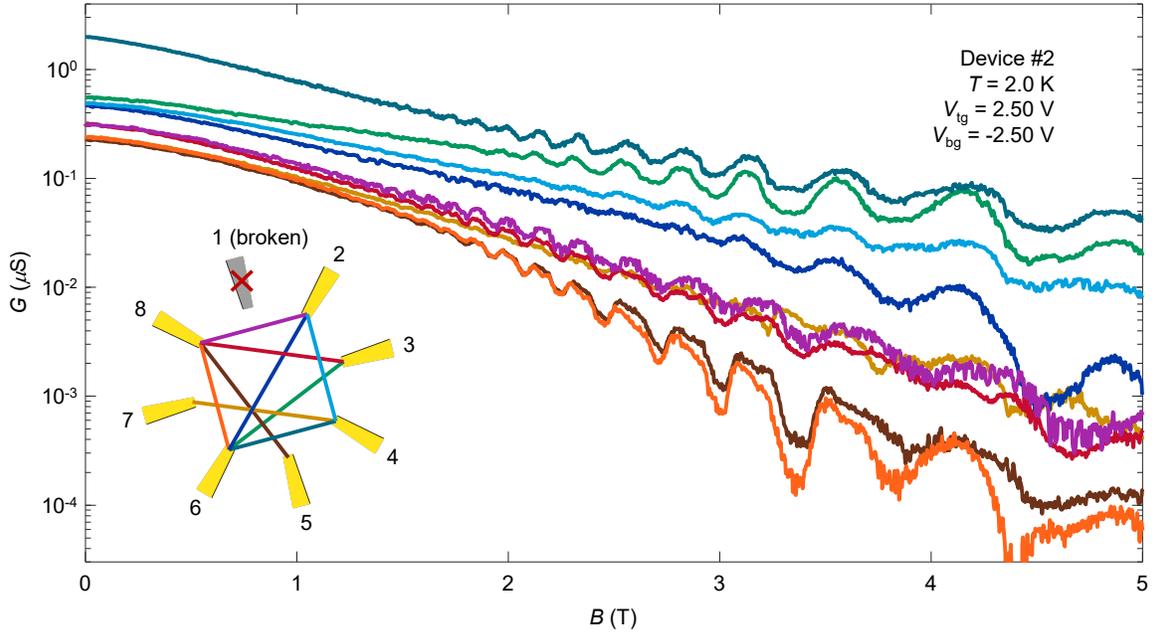

**Extended Data Fig. 5 | Quantum oscillations observed in different contact configurations in device #2.** The QOs are robustly observed between various contact pairs connected to the monolayer WTe$_2$ flake. In this device, 7 out of 8 contacts are working (contact #1 was broken during fabrication). Inset illustrates the corresponding contact pairs for each data curve. For each colored data curve, its source and drain contacts are indicated by connecting them with a solid line with the same color. The data curves were taken individually, i.e., all other contacts were floated when one pair of contacts were chosen to be measured. Data were recorded at 2 K.

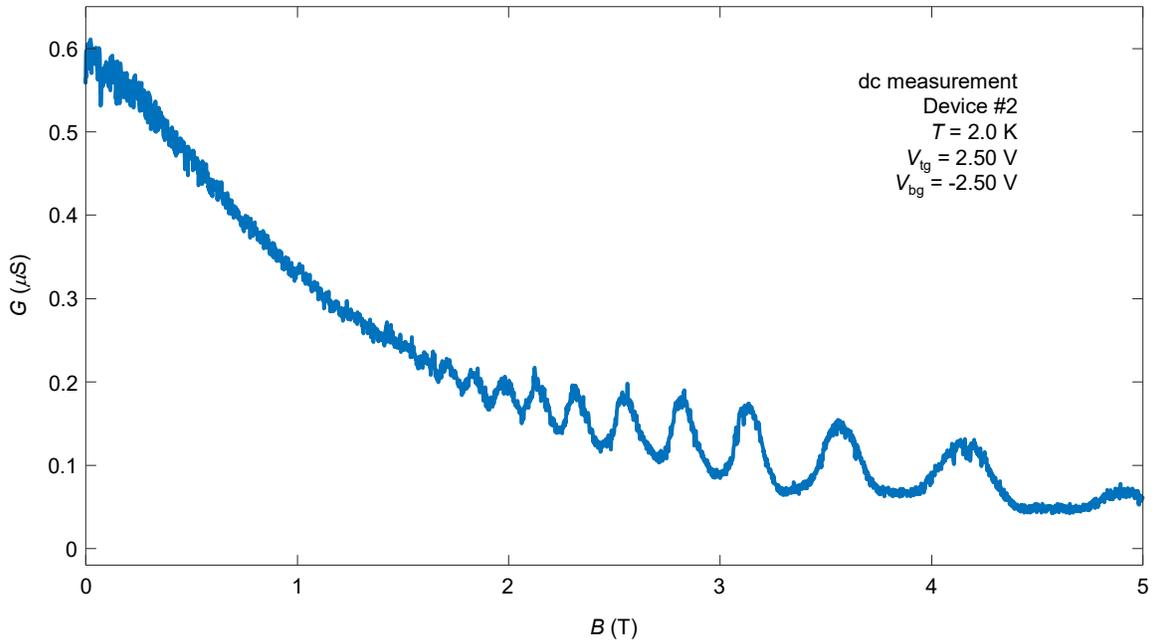

**Extended Data Fig. 6 | Quantum oscillations observed in dc measurements.** The two-probe dc conductance were measured by supplying a small dc excitation with an amplitude of 1.5 mV via a Keithley 2400 while monitoring the dc current. This strict dc measurement yields consistent results with the low frequency ac measurement used in the main text. Data were taken at 2 K from a pair of contacts in device #2.

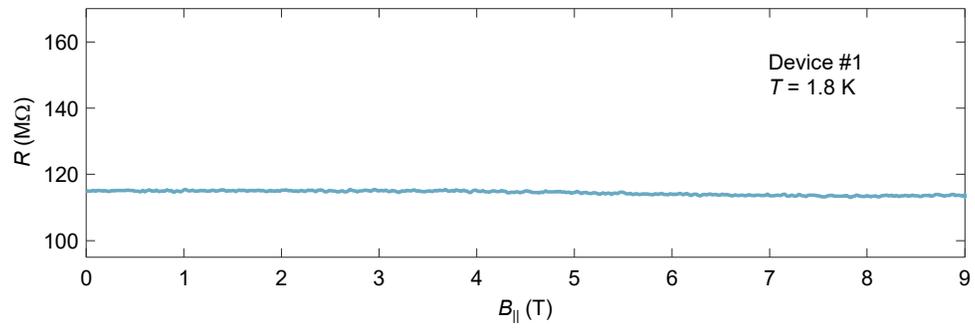

**Extended Data Fig. 7 | The effect of in-plane field.** The in-plane magnetoresistance of device #1 was taken using the same two contacts used in Fig. 2, in the insulating plateau at $T = 1.8$ K.

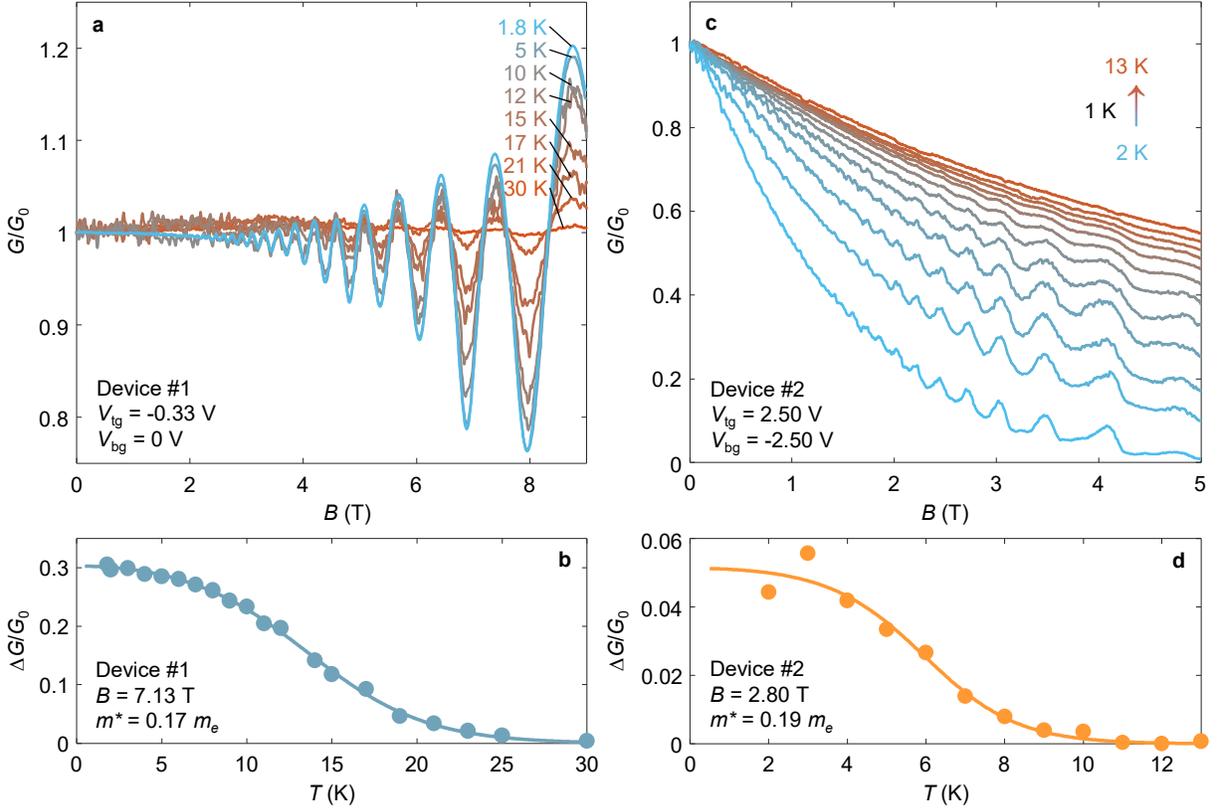

**Extended Data Fig. 8 | Additional analysis on the quantum oscillations in device #1 and #2. a,** Temperature dependence of conductance $G \equiv 1/R$ for device #1, normalized by its zero-field conductance. The same data set was used in Fig. 2, in which $R$ is plotted instead. **b,** Temperature-dependent oscillating amplitude at a selected field (7.13 T). The solid line is the fitting based on the modified LK formula described in the main text. Similar values are extracted for the fitting parameters compared to the analysis in Fig. 2. **c** and **d**, conductance data and LK fitting for device #2.

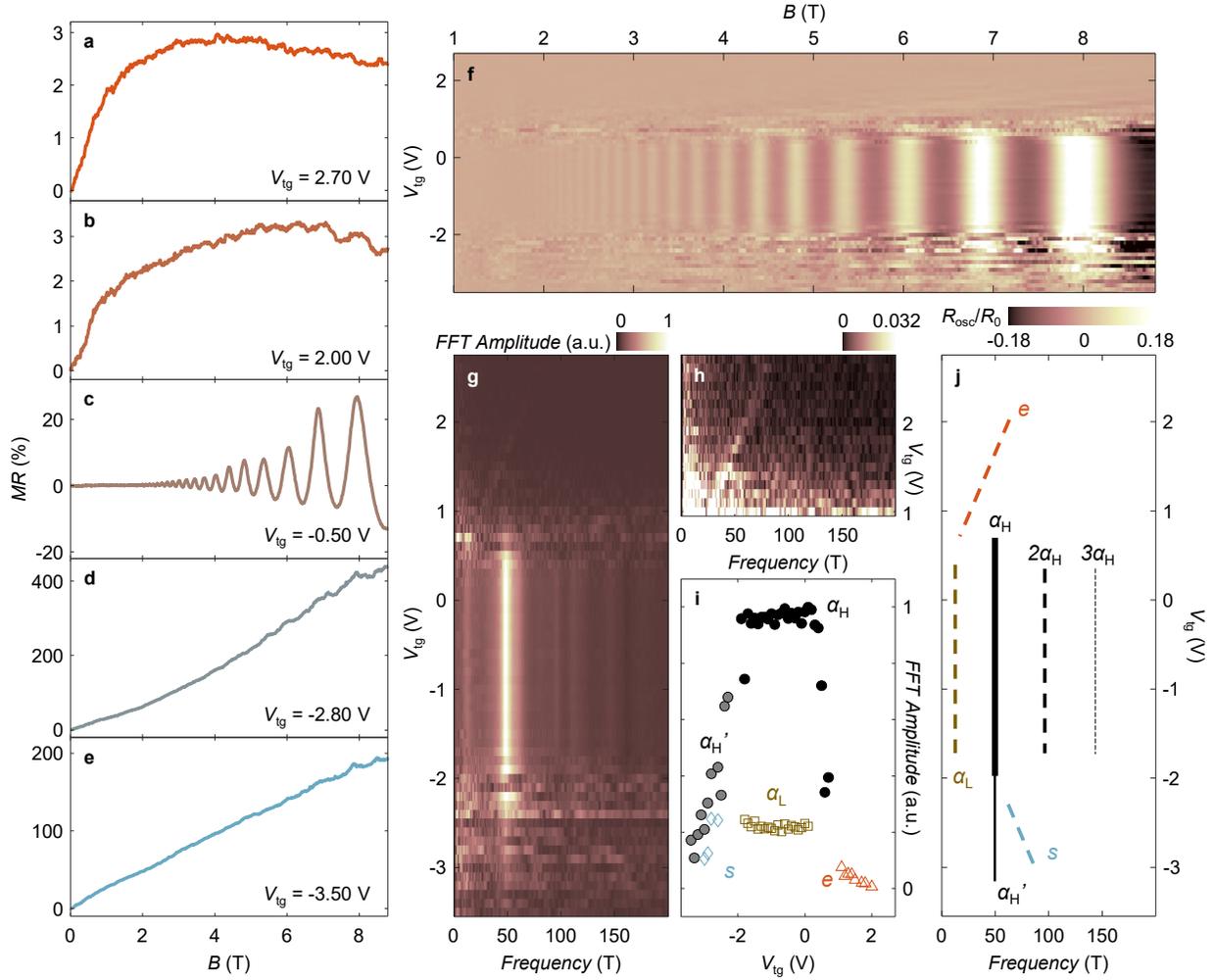

**Extended Data Fig. 9 | The effect of gating on the quantum oscillations in device #1. a-e,** Magnetoresistance curves taken at selected $V_{tg}$ ($V_{bg}$ = 0 V). **f,** Color map of the normalized oscillating component, $R_{osc}/R_0$, under varying $B$ and $V_{tg}$. **g,** FFT map of the same data. **h,** A zoom-in plot of the FFT map at $V_{tg} > 1$ V, highlighting the weak peak that features a gate-tunable frequency. **i,** Gate dependent FFT amplitudes of all visible peaks. **j,** A sketch summarizing the observed peaks. The solid black line ($\alpha_H$ peak) is the most pronounced peak and the focus of our discussion. Dashed color lines are weak features, whose amplitudes are shown in **i**. In the insulating regime, a lower frequency peak ($\alpha_L$) emerges together with the $\alpha_H$ peak. Towards the hole side, the $\alpha_H$ peak splits into two peaks ($\alpha_H'$ and $s$). Towards the electron side, a new peak ($e$) emerges from almost zero frequency with an amplitude that decreases monotonically. Coincident with the emergence of the $e$ peak, the amplitude of the $\alpha$ peak drops abruptly. Careful studies of these weak modes under higher magnetic fields and lower temperatures may provide further information to understand the system. Data recorded from a pair of contacts in device #1, at 1.8 K.

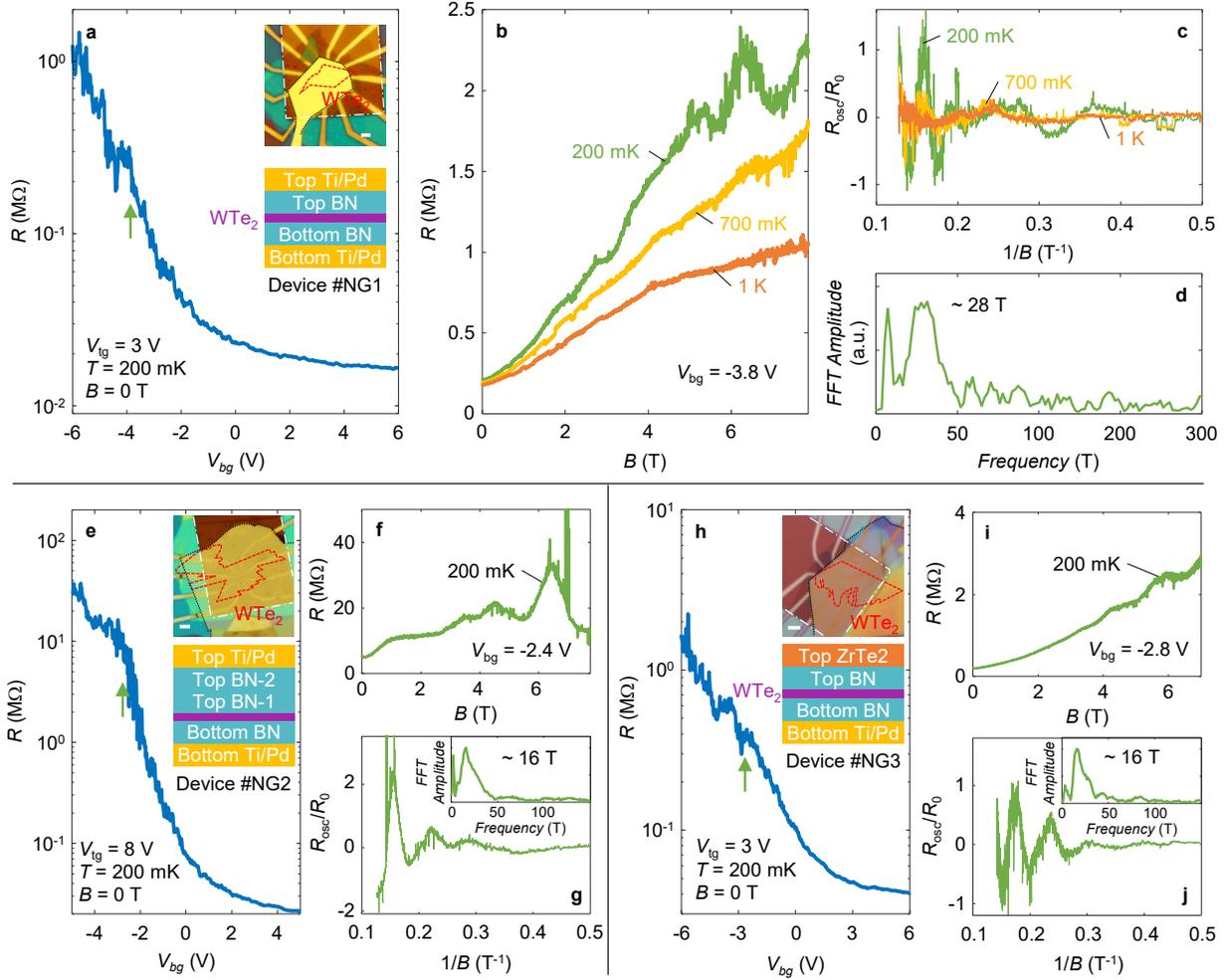

**Extended Data Fig. 10 | Magnetoresistance oscillations in non-graphite-gated devices. a,** Two-probe resistance ($R$) measured in device #NG1 as a function of the bottom gate voltage at a fixed top gate voltage. The device image and layer schematic are shown as inset. **b,** The magnetoresistance at a fixed $V_{bg}$, corresponding to a location indicated by the green arrow in **a.** The oscillations are visible at 200 mK but disappear at higher temperatures. **c,** The oscillating component of the magnetoresistance shown in **b** after background extraction. **d**, FFT of the 200 mK curve in **c**. **e,** Gate dependent $R$ and device information for another metal gated device #NG2. The device image and layer schematic are shown as inset. **f**, A magnetoresistance trace observed in the insulating regime (location indicated by the green curve in **e**). **g**, Its oscillation component together with the FFT. **h-j,** The same summary for device #NG3. The top gate is replaced by a thin $ZrTe_2$ flake. The $ZrTe_2$ thin flake is air-sensitive and its degradation could introduce inhomogeneities to the $WTe_2$ channel. We note that in all three devices, the observed sample resistances are lower than our typical graphite devices. Higher top gate voltages are also needed for the metal top-gated devices in order to perform measurements in the insulating regime (i.e., we expect significant contact resistance in the measured curves). Magnetoresistance oscillations only develop in the sub-Kelvin regime, indicating that freezing out the excess charge carrier is perhaps important to their appearance. All scale bars in the device images are 3 μm.

# Supplementary Information

# Landau Quantization and Highly Mobile Fermions in an Insulator

Pengjie Wang[1,#], Guo Yu[1,2,#], Yanyu Jia[1,#], Michael Onyszczak[1], F. Alexandre Cevallos[3], Shiming Lei[3], Sebastian Klemenz[3], Kenji Watanabe[4], Takashi Taniguchi[5], Robert J. Cava[3], Leslie M. Schoop[3], Sanfeng Wu[1,*]

[1] Department of Physics, Princeton University, Princeton, New Jersey 08544, USA
[2] Department of Electrical Engineering, Princeton University, Princeton, New Jersey 08544, USA
[3] Department of Chemistry, Princeton University, Princeton, New Jersey 08544, USA
[4] Research Center for Functional Materials, National Institute for Materials Science, 1-1 Namiki, Tsukuba 305-0044, Japan
[5] International Center for Materials Nanoarchitectonics, National Institute for Materials Science, 1-1 Namiki, Tsukuba 305-0044, Japan

[#] These authors contributed equally to this work
[*] Email: sanfengw@princeton.edu

**Contents**

## S1. Sample Fabrication Process

Below we describe our sample fabrication process in detail.

- **Summary of tools:**
    - **Dry transfer setup** in ambient conditions
    - **Fume hood:** dissolving polycarbonate (PC), metal lift-off, cleaning PMMA
    - **Furnace:** heat cleaning flakes and stacks
    - **Glovebox system**
        - Inert atmosphere: Ar
        - The concentration of $O_2$ and $H_2O$ are < 0.1 ppm.
        - A dry transfer setup installed inside.
    - **Reactive ion etching (RIE)** was performed by an Oxford PlasmaPro 80 RIE under a $CHF_3 + O_2$ plasma conditions.
    - **Electron beam lithography (EBL)** was performed by a Raith e-Line. The write field was 100 μm × 100 μm. Two-layer PMMA resists were applied as:
        a. 495PMMA A2, 2000rpm for 1 min, and baked at 180 °C for 7 mins
        b. 950PMMA A2, 3000rpm for 1 min, and baked at 180 °C for 3 mins
    - **Metal depositions** were performed by an e-beam evaporator (Angstrom Engineering Nexdep). The deposition began when the pressure reached 5E-7 torr, and the pressure was kept < 1E-6 torr during the deposition.
- **Complete steps for sample fabrications:**

Step 1-9 are for dual-graphite-gate devices. In step 10, we clarify the modified procedures for non-graphite-gate devices.

1. Exfoliation of graphite
    a. Exfoliate graphite on clean $Si/SiO_2$ wafers.
    b. Search for suitable graphite flakes for top and bottom gates (thickness: 5-10 nm).
    c. Image the flakes by atomic force microscope (AFM) to ensure cleanness.
2. Exfoliation of hBN
    a. Exfoliate hBN on clean $Si/SiO_2$ wafers.
    b. Heat clean at 450 °C for 6 hours in vacuum.
    c. Search for suitable hBN flakes for the top and bottom dielectrics (thickness: 5-12 nm) and the thin insulating hBN (thickness: 2-5 nm).
    d. AFM the flakes to ensure cleanness.
3. Creation of chips with align marks
    a. Dice undoped $Si/SiO_2$ wafer into appropriately sized pieces.
    b. Pattern align marks by EBL with 10kV acceleration voltage and 30 μm aperture size.
    c. Cold develop (0 °C) in isopropanol (IPA): deionized (DI) water (3:1 by volume) for 180s.
    d. Deposit Ti (5 nm) and Au (50 nm).

e. Lift off in acetone after a 1-hour bath, followed by sonication in another acetone bath for 5 mins. Once sonicated, transfer to dichloromethane for 1 hour and leave in a clean acetone bath for more than 3 hours.
4. Transfer bottom graphite and hBN (Extended Data Fig. 1b)
   a. Pick up graphite and hBN
   b. Transfer onto a clean chip patterned with align-marks.
   c. Remove PC with chloroform (1 h), dichloromethane (1 h), and chloroform (3 h).
   d. Heat clean at 400 °C for 6 hours in vacuum.
   e. AFM the stack and find a clean area (>7 µm) for contact electrodes.
5. Creation of contact and gate electrodes (Extended Data Fig. 1c)
   a. Pattern the main contact electrodes and gates by EBL with 10kV acceleration voltage and 30 µm aperture size.
   b. Pattern the outer bonding pads and connections by EBL with 10kV acceleration voltage and 60 µm aperture size (high current mode).
   c. Cold develop (0 °C) in IPA:DI water (3:1 by volume) for 180s.
   d. RIE for 10-16 s.
   e. Deposit Ti (3 nm) and Pd (17 nm).
   f. Lift off with acetone (1 h), dichloromethane (1 h), and acetone (>3 h).
   g. Tip clean electrodes by contact mode AFM (for a selected area < 10 µm).
6. Transfer thin hBN (Extended Data Fig. 1d)
   a. Pick up thin hBN.
   b. Align and cover thin hBN onto the contact electrodes at 50 °C.
   c. Gradually heat up to 150 °C for PDMS/PC detaching and to 180 °C for PC melted down.
   d. Remove the PC with chloroform (1 h), dichloromethane (0.5-1 h), and chloroform (3 h).
   e. AFM to check the cleanness and flatness after transfer.
7. Creation of holes for contacts (Extended Data Fig. 1e)
   a. Pattern the holes on the thin hBN by EBL with 10kV acceleration voltage and 30 µm aperture size. The work area is 100 µm, which is the same as in step 5a.
   b. Cold develop (0 °C) in IPA:DI water (3:1 by volume) for 180s.
   c. RIE the patterned holes depending on the thin hBN thickness.
   d. AFM the etched areas to ensure the thin hBN is fully etched and the contacts are exposed.
   e. Remove PMMA with acetone (1 h), dichloromethane (1 h), and acetone (>3 hours).
   f. Tip clean by contact mode AFM with an area no more than 10 µm.
   g. AFM to check the cleanness. (Extended Data Fig. 1e)
   h. Immediately move the chip with the bottom part into the glovebox.
8. Search and transfer WTe$_2$ and top gate (all were performed in the glove box)
   a. Exfoliate and search for WTe$_2$ in the glovebox.
   b. The monolayer WTe$_2$ is identified by optical image.
   c. Pick up top graphite by a clean PDMS/PC stamp. (Extended Data Fig. 1f)
   d. Stack graphite onto hBN at 60-70 °C.

e. Heat up to 95 °C.
  f. Cool down to around 65-70 °C and slowly pick up hBN for 15 mins. Bubble-free top parts are captured in Extended Data Fig. 1g.
  g. Pick up $WTe_2$ at 60 °C after heating the stamp up to 90 °C (Extended Data Fig. 1h).
  h. Carefully align the top parts to the bottom electrodes prepared in the last step and stack the graphite/hBN/$WTe_2$ stamp onto the electrodes. Leave the transferred chip in the glovebox for overnight. (Extended Data Fig. 1i)
9. Sample clean up and bonding
  a. Take out the sample from the glovebox and dissolve PC with chloroform (10 mins), dichloromethane (1 min), and chloroform (10 mins).
  b. Cut the chip to fit the chip carrier.
  c. Attach the chip onto the chip carrier by silver epoxy.
  d. Ground all the pins and wire bond Al wires to the pads of the contacts and gates.
  e. Transfer the chip carrier to fridge and cool down. The total time for step 9 is usually no more than 40 mins.
10. For the non-graphite-gated devices (#NG1-3), we modified the steps as following:
  a. Bottom gate preparation: step 4 is modified to prepattern and deposit the bottom metal gate with Ti/Pd (3nm/8-17nm), then follow with either AFM tip cleaning or annealing, and hBN transfer.
  b. Top gate preparations are different in the three samples:
    i. For #NG1, at step 8, we only picked up hBN/$WTe_2$. After step 9a, the sample was tip-cleaned by contact mode AFM, then patterned by EBL and deposited with Ti/Pd (5 nm/50 nm). After lift-off, we followed the steps 9b-e.
    ii. For #NG2, we selected a chip with exfoliated hBN in step 2, and evaporated with Ti/Pd (3nm/13nm) (labeled as hBN-2 in Extended Data Fig. 10e). After step 7, we picked up the hBN-2, and stacked it onto another hBN (labeled as hBN-1 in Extended Data Fig. 10e). After dissolving the PC, the stack was moved into the glovebox. In step 8, we picked up graphite, metal/hBN-2/hBN-1, and $WTe_2$, and then stacked the stamp onto the electrodes. The use of graphite is to merely connect the metal top gate on top of hBN-2 to the outside gate electrodes (for wire bonding). The graphite covers no regimes of the monolayer $WTe_2$. The cleanup and bonding process is the same as step 9.
    iii. For #NG3, we bonded the chip (step 9b-d) after AFM tip cleaning (step 7g). We then exfoliated and searched the thin $ZrTe_2$ flake in the glovebox and followed step 8 by replacing the graphite with $ZrTe_2$. When picking up the $ZrTe_2$, the stamp was heated up to 140°C in order to pick up the flake. Step 8h was done with the chip bonded on a chip carrier. The chip carrier was then moved to the fridge and cooled down after the transfer. The PC layer was kept on the top of the sample as protection for $ZrTe_2$ ($ZrTe_2$ thin flake is air-sensitive).

## S2. Summary of Device Parameters

|  | Device #1 | Device #2 | Device #3 |
|---|---|---|---|
| Top Graphite | ~ 4 nm | ~ 10 nm | ~ 7 nm |
| Top BN | 5.6 nm | 12.1 nm | 11.0 nm |
| Thin BN | 3.8 nm | 3.5 nm | 4.8 nm |
| Bottom BN | 6.1 nm | 8.6 nm | 9.2 nm |
| Bottom Graphite | 4.9 nm | 8.7 nm | 3.1 nm |
| Bulk $WTe_2$ | Vapor growth | Vapor growth | Flux growth |

|  | Device #NG1 | Device #NG2 | Device #NG3 |
|---|---|---|---|
| Top Gate | Ti/Pd 55.0 nm | Ti/Pd 16.0 nm | $ZrTe_2$ ~ 40 nm |
| Top BN | 7.4 nm | 16.5 nm + 12.7 nm | 12.0 nm |
| Thin BN | 3.0 nm | 1.8 nm | 1.7 nm |
| Bottom BN | 9.2 nm | 18.4 nm | 13.1 nm |
| Bottom Gate | Ti/Pd 13.5 nm | Ti/Pd 17.5 nm | Ti/Pd 11.0 nm |
| Bulk $WTe_2$ | Vapor growth | Flux growth | Flux growth |

## S3. Fitting Procedure Using Modified Lifshitz-Kosevich (LK) Formula

To fit the temperature dependence data shown in Fig. 2d and Extended Data Fig. 7 (in the quantum oscillation regime), we use a modified LK formula, written as:

$$\Delta R/R_0 \sim \Delta G/G_0 \sim \left[\frac{1}{\gamma(B)} + \frac{\sinh\left(\frac{\beta T m^*}{B}\right)}{R_D(B)\left(\frac{\beta T m^*}{B}\right)}\right]^{-1}.$$

Here $\beta = 2\pi^2 k_B/e\hbar$, $\hbar$ is the reduced Plank constant, $k_B$ is the Boltzmann constant, $e$ is the elementary charge, $m^*$ is the effective mass and $R_D(B) \propto e^{-\frac{D}{B}}$ is the Dingle damping term where $D = \beta m^* T_D = \frac{\pi m^*}{\tau} = \frac{\pi}{\mu}$ ($T_D$ is the Dingle temperature, $\tau$ is the scattering time and $\mu$ is the mobility). $\gamma(B)$ is a new term we introduce to describe the saturating behavior at low $T$. For each oscillation period, the peak-to-valley difference $\Delta R$ ($\Delta G$) is normalized by its zero-field resistance $R_0$ ($G_0$). The corresponding value for $B$ is determined by $B = 2[1/B_{peak}+1/B_{valley}]^{-1}$. In Fig. 2, $\Delta R/R_0$ in the range of $B > 4.8$ T has well-defined temperature dependence, and hence are used in our fitting. We plot the extracted fitting parameters in Fig. S1. The fitting results on the same data set based on $\Delta G/G_0$ is shown in Extended Data Fig. 7a and b. The data and results for device #2 are included in Extended Data Fig. 7c and d. Consistent fitting results are obtained in all data.

Figure S1a plots the extracted effective mass $m^*$ as a function of the field $B$ based on the fittings in Fig. 2 ($\Delta R/R_0$). Figure S1b shows the Dingle damping term in a log $R_D$ - $1/B$ plot. The solid line displays a linear fitting, which yields $D = 29.5$ T. The corresponding mobility is then $\mu$

~1,100 cm$^2$V$^{-1}$s$^{-1}$. Figure S1c plots $\gamma(B)$ on a log-log scale and its best linear fit. The data indicates a power-law relation $\gamma(B) \sim B^v$ where the exponent $v \sim 3.0$. Fittings based on $\Delta G/G_0$ yields similar results ($\mu \sim 1{,}100$ cm$^2$V$^{-1}$s$^{-1}$ and $v \sim 2.7$).

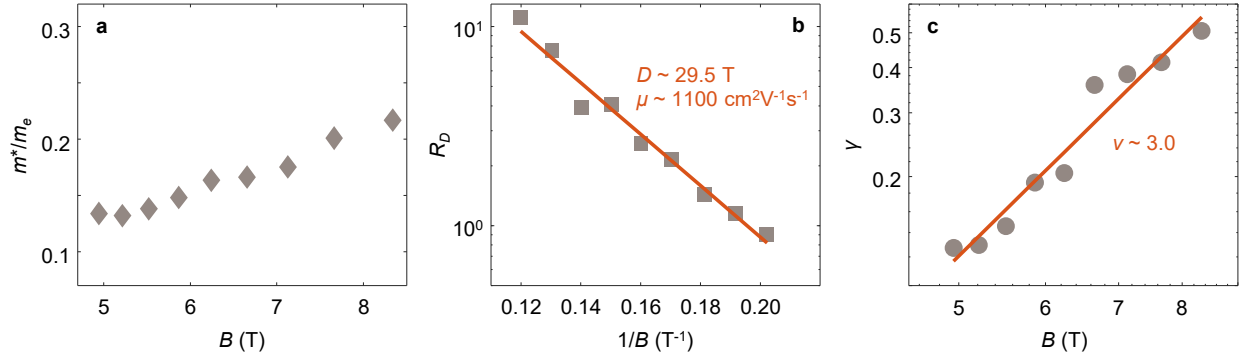

**Figure S1 | Fitting parameters in the modified LK formula. a,** Effective mass as a function of magnetic field. **b,** $R_D$ damping term as a function of $1/B$. **c,** $\gamma$ term as a function of $B$ in a log-log scale.